\documentclass{elsart5}

 \usepackage{epsfig}

\usepackage{amssymb}

\begin{document}

\begin{frontmatter}

\title{The ALPS project release 1.3: \\ open source software for strongly correlated systems}

\author[affeth]{A.F. Albuquerque}
\author[afftls]{F. Alet}
\author[affeth]{P. Corboz}
\author[affeth,affmin]{P. Dayal}
\author[affq]{A. Feiguin}
\author[affgoet]{S. Fuchs}
\author[affeth]{L. Gamper}
\author[affeth]{E. Gull}
\author[affhkg1,affhkg2]{S. G\"urtler}
\author[affgoet]{A. Honecker}
\author[afftokyo2]{R. Igarashi}
\author[affeth]{M. K\"orner}
\author[affru]{A. Kozhevnikov}
\author[affepfl]{A. L\"auchli}
\author[affmar,affstut]{S.R. Manmana}
\author[affeth]{M. Matsumoto}
\author[affrwth]{I.P.~McCulloch}
\author[affgraz]{F. Michel}
\author[affmar]{R.M.~Noack}
\author[affpol]{G. Paw{\l}owski}
\author[affeth]{L. Pollet}
\author[affgoet]{T. Pruschke}
\author[affrwth]{U.~Schollw\"ock}
\author[afftokyo]{S. Todo}
\author[affq]{S. Trebst}
\author[affeth,cor1]{M.~Troyer}
\ead{troyer@comp-phys.org}
\corauth[cor1]{}
\author[affcol]{P.~Werner}
\author[affstut]{S.~Wessel}  
\\ (for the ALPS collaboration)

\address[affeth]{Theoretische Physik, ETH Z\"urich, 8093 Z\"urich, Switzerland}
\address[afftls]{Laboratoire de Physique Th\'eorique, UMR CNRS 5152,
Universit\'e Paul Sabatier, 31062 Toulouse, France}
\address[affmin]{University of Minnesota, Department of Computer Science and Engineering, 200 Union Street S.E. Minneapolis, MN 55455, USA}
\address[affq]{Microsoft Research, Station Q, University of California,
Santa Barbara, CA 93106, USA}
\address[affgoet]{Institut f\"ur Theoretische Physik, Universit\"at G\"ottingen, D-37077 G\"ottingen, Germany}
\address[affhkg1]{Centre for Theoretical \& Computational Physics, The University of Hong Kong, Hong Kong, China}
\address[affhkg2]{Department of Physics, The University of Hong Kong, Hong Kong, China}
\address[afftokyo2]{Department of Physics, University of Tokyo, 113-0033 Tokyo, Japan}
\address[affru]{Institute of Metal Physics, Russian Academy of Sciences -- Ural Division,
620219 Ekaterinburg GSP-170, Russia}
\address[affepfl]{Institut Romand de Recherche Num\'erique en Physique des Mat\'eriaux (IRRMA),
CH-1015 Lausanne, Switzerland}
\address[affmar]{AG Vielteilchennumerik, Fachbereich Physik, Philipps-Universit\"at Marburg, D-35032 Marburg, Germany}
\address[affstut]{Institut f\"ur Theoretische Physik III, Universit\"at Stuttgart, Pfaffenwaldring 57, D-70550 Stuttgart, Germany}
\address[affrwth]{ Institut f\"ur Theoretische Physik C, RWTH Aachen, D-52056 Aachen, Germany}
\address[affgraz]{ Institut f\"ur Theoretische Physik, Technische Universit\"at Graz, Petersgasse 16, A-8010 Graz, Austria}
\address[affpol]{Institute of Physics, A. Mickiewicz University, ul. Umultowska 85, 61-614 Poznan, Poland
}
\address[afftokyo]{Department of Applied Physics, University of Tokyo, 113-8656 Tokyo, Japan}
\address[affcol]{Department of Physics, Columbia University, 538 West 120th Street, New York,
NY 10027, USA}

%\received{12 June 2005}
%\revised{13 June 2005}
%\accepted{14 June 2005}
%use optional labels to link authors explicitly to addresses:

%\author{}
%\address{}

\begin{abstract}
We present release 1.3 of the ALPS (Algorithms and Libraries for Physics Simulations)
project, an international open source software project to develop
libraries and application programs for the simulation of strongly
correlated quantum lattice models such as quantum magnets, lattice
bosons, and strongly correlated fermion systems. Development is
centered on common XML and binary data formats, on libraries to
simplify and speed up code development, and on full-featured
simulation programs. The programs enable non-experts to start carrying
out numerical simulations by providing basic implementations of the
important algorithms for quantum lattice models: classical and quantum
Monte Carlo (QMC) using non-local updates, extended ensemble
simulations, exact and full diagonalization (ED), as well as the
density matrix renormalization group (DMRG). Changes in the new release include a DMRG program for interacting models, support for translation symmetries in the diagonalization programs, the ability to define custom measurement operators, and support for inhomogeneous systems, such as lattice models with traps. 
The software is available
from our web server at {\tt http://alps.comp-phys.org/}.\end{abstract}

%%%%%%%%%use  the \KEY command at the begin of keyword text%%%%%%%%%
\begin{keyword}
\PACS 02.70.-c\sep 75.40.Mg
\KEY quantum lattice model \sep open source software \sep C++, Monte Carlo \sep quantum Monte Carlo \sep density matrix renormalization group \sep DMRG \sep exact diagonalization
\end{keyword}
%Please supply one or more relevant PACS-1996 classification codes 
%(http://publish.aps.org/PACS/96pacs.htmland) and about 5 keywords 
%of your own choice for indexing purposes. 
%You can see a list of already used keywords for JMMM at 
%http://authors.elsevier.com/JournalDetail.html?PubID=505704&Precis=KIND

\received{\today}
\accepted{}
\end{frontmatter}

\section{Introduction}
\label{}

In this paper we present release 1.3 of the ALPS project  (Algorithms and Libraries for Physics Simulations), an open source software development project for strongly correlated lattice models. We will present a short overview  and focus on new features compared to the previous releases \cite{ALPS1.2,alps}.

Quantum fluctuations and competing interactions in quantum many body
systems lead to unusual and exciting properties of strongly correlated
materials such as quantum magnetism \cite{QMbook}, high temperature
superconductivity \cite{BednorzMueller}, heavy fermion
behavior \cite{HeavyFermion}, and topological quantum order \cite{Wen}.
The same strong interactions make accurate analytical treatments hard and 
direct numerical simulations are essential to increase our understanding of the unusual
properties of these systems. 

The last decade has seen tremendous progress in the development of
algorithms.  Speedups of many orders of magnitude have been 
achieved \cite{cluster,QWL,DMRG,DMRGreview,EDDMRG,Werner06}. These
advances come at the cost of substantially increased algorithmic
complexity and challenge the current model of program development in
this research field. In contrast to other research areas, in which
large ``community codes'' are being used, the field of strongly
correlated systems has so far been based mostly on single codes developed by
individual researchers for particular projects. While the simple
algorithms used a decade ago could be easily programmed by a beginning
graduate student in a matter of weeks, it now takes substantially
longer to master and implement the new algorithms.  Thus, their use
has increasingly become restricted to a small number of experts.

The ALPS project aims to
overcome the problems posed by the growing complexity of algorithms
and the specialization of researchers onto single algorithms through
an open-source software development initiative. In release 1.3 new features desired by ALPS users, especially experimentalists, have been added, as well as a DMRG program for interacting problems.

To quote from the previous publication \cite{ALPS1.2}, the goals of the ALPS project are to
provide:
\begin{itemize}
\item {\bf standardized file formats} to simplify exchange,
distribution and archiving of simulation results and to achieve
interoperability between codes.
\item {\bf generic and optimized libraries} for common aspects of
simulations of quantum and classical lattice models, to simplify code
development.
\item a set of {\bf applications} covering the major algorithms.
\item{\bf license} conditions that encourage researchers to contribute
to the ALPS project by gaining scientific credit for use of their
work.
\item {\bf outreach} through a web page \cite{alps}, mailing lists and
workshops to distribute the results and to educate researchers both
about the algorithms and the use of the applications.
\item {\bf improved reproducibility} of numerical results by
publishing source codes used to obtain published results.
\item an {\bf archive} for simulation results.
\end{itemize}

The ready-to-use applications are useful both for 
{\it theoreticians} who want to test theoretical ideas about quantum
lattice models and to explore their properties, as well as for 
{\it experimentalists} trying to fit experimental data to theoretical
models to obtain information about the microscopic properties of
materials.

 In the following, we present a quick review of these
 aspects of the ALPS project, focusing on new features in release 1.3.
\section{File formats}

The most basic part of the ALPS project is the definition of
common standardized file formats suitable for a wide range of
applications. Standardized file formats enable the exchange of data
between applications, allow the development of common evaluation
tools, simplify the application of more than one algorithm to a given
model, and are a prerequisite for the storage of simulation data in a
common archive. The ISO
standard XML \cite{xml} was chosen for the specification of these formats
because it has become
the main text-based data format on the internet and because it is
supported by a large and growing number of tools. 

We have designed a number of XML  schemas \cite{xmlschema} to describe
\begin{itemize}
\item the input of simulation parameters,
\item the lattices,
\item quantum lattice models,   
\item and the output of results.
\end{itemize}

\begin{figure}
\begin{small}
\begin{verbatim}
<LATTICEGRAPH name = "depleted inhomogeneous square lattice">
  <FINITELATTICE>
    <LATTICE dimension="2"/>  
    <EXTENT dimension="1" size="L"/>
    <EXTENT dimension="2" size="L"/>
    <BOUNDARY type="periodic"/>  
  </FINITELATTICE>
  <UNITCELL>
    <VERTEX/>
    <EDGE>
      <SOURCE vertex="1" offset="0 0"/>
      <TARGET vertex="1" offset="0 1"/>
    </EDGE>
    <EDGE>
      <SOURCE vertex="1" offset="0 0"/>
      <TARGET vertex="1" offset="1 0"/>
    </EDGE>
  </UNITCELL> 
  <INHOMOGENEOUS><VERTEX/></INHOMOGENEOUS>}
  <DEPLETION>
    <VERTEX probability="DEPLETION" seed="DEPLETION_SEED"/>
  </DEPLETION>
</LATTICEGRAPH>
\end{verbatim}

\end{small}
\caption{The definition of a square lattice with one site (vertex) per
unit cell and bonds (edges) only to nearest neighbors. First the
dimension, extent, and boundary conditions of the Bravais lattice are
described in the {\tt <FINITELATTICE>} element, then the unit cell
including the bonds in the lattice is defined. New features in release 1.3. include inhomogeneity and depletion.}
\label{fig:lattice}
\end{figure}

\begin{figure}
\begin{small}
\begin{verbatim}
<BASIS name="boson">
<SITEBASIS>
  <PARAMETER name="Nmax" default="infinity"/>
  <QUANTUMNUMBER name="N" min="0" max="Nmax"/>
  <OPERATOR name="bdag" matrixelement="sqrt(N+1)">
    <CHANGE quantumnumber="N" change="1"/>
  </OPERATOR>
  <OPERATOR name="b" matrixelement="sqrt(N)">
    <CHANGE quantumnumber="N" change="-1"/>
  </OPERATOR>
  <OPERATOR name="n" matrixelement="N"/>
</SITEBASIS>
</BASIS>

<SITEOPERATOR name="double_occupancy" site="x">
  n(x)*(n(x)-1)/2
</SITEOPERATOR>

<BONDOPERATOR name="boson_hop" source="x" target="y">
  bdag(x)*b(y)+bdag(y)*b(x)
</BONDOPERATOR>

<HAMILTONIAN name="trapped boson Hubbard">
  <PARAMETER name="mu" default="0"/>
  <PARAMETER name="t" default="1"/>
  <PARAMETER name="U" default="0"/>
  <PARAMETER name="K" default="0"/>
  <BASIS ref="boson"/>
  <SITETERM site="i">
    -mu*n(i) + U*double_occupancy(i) + K/2*((x-L/2)^2 + (y-L/2)^2)
  </SITETERM> 
  <BONDTERM source="i" target="j">
    -t*boson_hop(i,j)
  </BONDTERM>
</HAMILTONIAN>
\end{verbatim}
\end{small}
\caption{The definition of a bosonic Hubbard Hamiltonian (1) in a harmonic trap.  After
  describing the local basis for each site and operators acting on it, the first new feature in release 1.3 is the ability to define composite site and bond operators, such as the double occupancy and bosonic hopping terms, which are then used in the definition of the Hamiltonian.}
\label{fig:model}
\end{figure}

A detailed specification of the formats is given on
our web pages \cite{alps,xmlschema}. As examples we show a lattice and a model definition in figures \ref{fig:lattice} and
\ref{fig:model}.   Any of the ALPS applications can be run by providing an input file, specifying the simulation parameters,
together with lattice and model definitions (figures \ref{fig:lattice}
and \ref{fig:model}) to that application (provided the application
supports that type of model). 
Standardized formats that extend across all applications reduce the
learning curve for using the applications and allow common tools to be
used to analyze the data.

\subsection{New features}
\subsubsection{Lattice definitions}
New features in the lattice definitions in release 1.3 include the possibility to specify inhomogeneity and depletion. While in a regular lattice, the model Hamiltonian is the same for all bonds or sites with the same type, in an inhomogeneous lattice the couplings in the model can be different for each bond or site. Examples are disordered systems, with randomly chosen couplings such as spin glasses, or systems in spatially varying trapping potentials, such as optical lattices in harmonic traps.
Inhomogeneities can be specified for vertices (sites) and edges (bonds), either for all (such as for all vertices in figure \ref{fig:lattice}), or only for one type of vertex or edge.

The other new feature is depletion of a lattice, where a fraction of sites or bonds is randomly removed from a lattice. Currently only site depletion is implemented, but additional types of depletion can be added easily if  the need arises \cite{contact}. In the example in figure  \ref{fig:lattice}, the fraction of depleted sites is specified to be passed in the {\tt DEPLETION} input parameter, and the random number generator seed {\tt DEPLETION\_SEED} can be changed to give different random realizations of the depletion pattern.

\subsubsection{Model definitions}
Figure \ref{fig:model} shows the definition of the Hamiltonian of a bosonic Hubbard model in a harmonic trap
 \begin{eqnarray}
 H&=&-\mu\sum_in_i+\frac{U}{2}\sum_in_i(n_i-1)   \\
    && -t \sum_{\langle
    i,j\rangle} \left(b_i^\dag b_j + b_j^\dag b_i\right) + \frac{K}{2}\sum_i(x_i^2+y_i^2). \nonumber
  \end{eqnarray}
  The first new feature of release 1.3, shown in this example, is the ability to specify composite site and bond operators, such as the {\tt double\_occupancy} site term or the {\tt boson\_hop} bond term. These site and bond operators can be used in the specification of the model and of measurements.
  
  The other new feature, also shown in the example, is the ability, in conjunction with the specification of an inhomogeneous lattice,  to use site-dependent couplings, such as the harmonic trapping potential $K$ which depends on the $x$ and $y$ coordinates of the site. This feature has been used for the simulation of cold bosonic gases in optical lattices \cite{boson}.
  
\subsection{Future plans}
The immediate goal for release 1.4 will be to add multi-site interaction terms and measurements, such as ring exchange terms.

With the experience gained in using common file formats for several years across a number of different applications, we next plan to design more widely used common formats, such as standard formats for results of Monte Carlo simulations that have been defined at a workshop in September 2006.
% and which will be published soon. 
%We expect to set more such standards in the future, in collaboration with other groups.

\section{Libraries}
The ALPS libraries are the foundation of all the ALPS applications,
providing functionality common to all of them, such as file I/O, expression evaluation, parallel job scheduling, or the statistical analysis of Monte Carlo measurements.
These libraries make full use of object-oriented and generic
programming techniques \cite{CE}, which allows them to be very flexible without
losing any performance compared to FORTRAN programs.

\subsection{New features}

The new library features are mainly hidden from the application users, providing support for the new features in the input files, discussed above, or providing support for new features in the applications, such as translational symmetries in the diagonalization codes, and the ability to define custom measurements, which will be discussed below.

One new feature that should be mentioned explicitly is the inclusion of random number generators in the expression library. Expressions for couplings and other parameters can make use of new functions {\tt random()}, {\tt normal\_random()}, and {\tt integer\_random(n)} to obtain uniform random numbers in the interval $[0,1)$, normally distributed random numbers or integer random numbers between $0$ and $n$. This feature is most useful in conjunction with inhomogeneous lattices to define disordered models.

Another change to be mentioned is that since the parsing of command line options to the application is now handled by the Boost program options library \cite{boost}, multi-letter command line options now need to be preceded by two hyphens as in {\tt --Tmin}, instead of the previous usage  {\tt -Tmin}.

\subsection{Future plans}

Code from the ALPS libraries has recently been used in the development of Boost \cite{boost} libraries. These include optimizations to the Boost.Serialization library, new features for the Boost.Random library, a proposed Boost.MPI library and a Boost.Accumulator library for statistical estimates. Once accepted in the peer review process, and published in a future Boost release, the ALPS codes will be changed to use these Boost libraries. The ALPS libraries will then contain less general purpose codes (which will have been moved to Boost) and will be more focused  on domain-specific libraries for the simulation of quantum models.

\section{Applications}
\label{sec:applications}
In addition to common libraries, the ALPS project includes a number of ready to use applications implementing the most important unbiased
algorithms for quantum many body systems. The applications all
share the same file formats, simplifying their use, reducing the
learning curve, and enabling the easy investigation of a model with
more than one method. Tutorials on the use of the applications are
included with the sources that can be found on the ALPS web
page \cite{alps}.

\subsection{Exact diagonalization}
The existing exact diagonalization \cite{ED,EDDMRG} programs  {{\tt sparsediag} and {\tt fulldiag} have been substantially improved. {\tt sparsediag} calculates the ground state
and low lying excited states of quantum lattice models using the
Lanczos \cite{lanczos} algorithm, while {\tt fulldiag} calculates the complete
spectrum of quantum lattice models and from it all thermodynamic
properties.

The first major improvement is the use of translation symmetry to reduce the Hilbert space dimension. This speeds up the calculations and allows larger systems to be calculated. In addition, by calculating the energy eigenvalues separately for each momentum, the momentum-resolved excitation spectrum can be calculated.

The second new feature is the possibility to define custom measurements of averaged or local site- and bond operators, as well as arbitrary 2-point correlation functions. As an example, for the inhomogeneous bosonic Hubbard model of figure \ref{fig:model}, one can specify the measurement of the average double occupancy, local density, density correlations, and Green's function by defining the following input parameters, which use the definitions of site and bond operators provided with the model:

\begin{verbatim}
MEASURE_AVERAGE[Double] = double_occupancy
MEASURE_LOCAL[Local density] = n
MEASURE_CORRELATION[Density correlation] = n
MEASURE_CORRELATION[Green function] = "bdag:b"
MEASURE_STRUCTURE_FACTOR[Spin Str Fact] = Sz
\end{verbatim}

In the future we plan to support point group symmetries and to add the calculation of dynamical correlation functions.

\subsection{Classical Monte Carlo}

The Monte Carlo program for classical magnets employing local and cluster
updates \cite{SW} has been extended according to wishes of several users. The most notable new features are the possibility to specify single ion anisotropies and anisotropic interactions.

\subsection{Quantum Monte Carlo}

The quantum Monte Carlo programs have also been updated in release 1.3.

The  ``looper'' program using the loop cluster
algorithm \cite{cluster,looper} in a stochastic series
expansion \cite{sse} (SSE) and path-integral representation for quantum
magnets has been replaced with a new version, offering 40-80\% speed increase and supporting longitudinal and transverse magnetic fields as well as single ion anisotropies.

The directed loop QMC program \cite{directed} in an SSE
representation for bosonic and magnetic quantum lattice models has been extended to offer also the ``Locally optimal worm algorithm" (LOWA) \cite{LOWA} updates. 

The worm algorithm \cite{worm} program has been updated to allow nonlocal interactions \cite{previousworm}.
  
All QMC program have been extended by new measurements, including correlation functions, structure factors, and stiffness or superfluid density. In addition, the same custom measurements as discussed above for exact diagonalization can be used in these codes, as long as the measurements are diagonal in the chosen basis. The looper program additionally measures local susceptibilities and transverse magnetization.

\subsection{Density Matrix Renormalization Group}

In addition to the DMRG \cite{DMRG,DMRGreview,EDDMRG} program for {\em noninteracting}
particles  (an adaptation of the program 
in Refs.\ \cite{dmrgbook,noninteracting} to the ALPS libraries), release 1.3 contains a new DMRG program for interacting many body systems, with the ability to calculate local quantities and two-point correlation functions.

\subsection{Tools}

Release 1.3 adds a database tool, to archive a collection of simulation results, and to allow fast and efficient extraction and evaluation of results. The tool, based on the SQlite \cite{sqlite} database is easy to use, based on a simple flat file database and does not require to set up a complex database application.

\section{License}
The ALPS software is distributed under the ALPS library \cite{liblicense} and
application \cite{applicationlicense} licenses, which have remained unchanged. The use of the
 codes is free for noncommercial applications but carries a citation requirement to publications, such as the current one, describing the
libraries and codes. In addition, besides the required citation of the publication describing the ALPS application, users are encouraged to cite the publication describing the algorithm used in the applications.

\section{Distribution, outreach development}
The ALPS web page \cite{alps} is the central information
repository of the ALPS collaboration.  It makes source codes of the
ALPS libraries and applications available, provides access to the
documentation, and distributes information about the mailing
lists, new releases, and workshops.

For this release we have replaced the static web page by a Wiki system \cite{wiki}, which enables developers and users to easily and quickly add new contents and improve the instructions and tutorials on the web page.

Another new feature is cdALPS \cite{cdalps} by G. Pawlowski, a bootable Knoppix Linux ALPS CD, which allows to run the ALPS tutorials and simple ALPS codes on a PC without installing the full ALPS software.

Development of the ALPS libraries and applications is coordinated
through both mailing lists and semi-annual developer workshops. Interested 
researchers are invited to join the mailing list and participate in the workshops.

\section{Future plans}
\subsection{New application}
The following new applications are planned for future releases:
\begin{itemize}
\item A series expansion \cite{series} code for both perturbation
series and high temperature series in release 1.4
\item Also in release 1.4, the addition of generalized sampling algorithms, such as the quantum version \cite{QWL} of the
Wang-Landau algorithm \cite{WangLandau} and optimized ensemble methods \cite{opt} to the quantum Monte Carlo algorithms to obtain thermodynamic quantities over large temperature ranges.
\item Implementation of parallel tempering \cite{pt} for the classical and quantum Monte Carlo algorithms including feedback-optimized temperature sets \cite{optpt}.
\item A QMC algorithm in the valence bond representation for quantum spin systems \cite{sandvik}
\item A continuum QMC program for bosons, implementing the continuous space worm algorithm \cite{cworm}.
\end{itemize}

\subsection{Realistic model: interface to band structure codes}

Also planned for release 1.4 is an interface of the ALPS applications to band structure codes. Instead of running simulations on phenomenological toy models, such as a simple Heisenberg model with only a few different couplings, it will be possible to run the ALPS applications on a complex, realistic model obtained from ab-initio band structure calculations. In the past this was done manually, e.g. in reference \cite{vanadates}, the exchange couplings determined in an LDA+U calculation \cite{ldau} using the Lichtenstein method \cite{lichtenstein} were used to manually prepare input files for QMC simulations using the loop algorithm \cite{cluster}.

\begin{figure}
\begin{center}
\includegraphics[width=8cm]{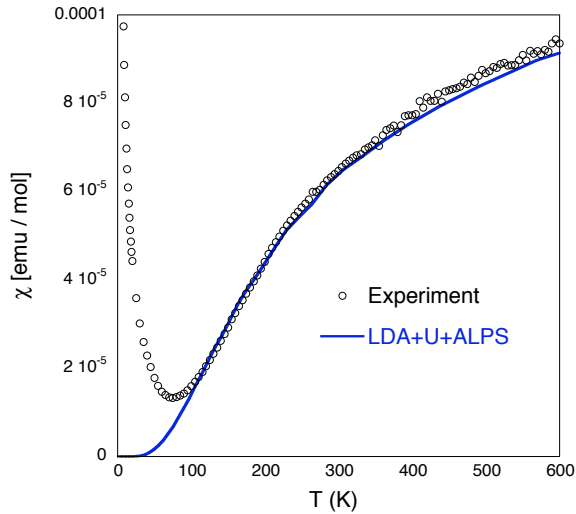}
\end{center}
\caption{Comparison of experimental measurements on the spin ladder compound SrCu$_2$O$_3$ \cite{azuma} with automated ab-initio simulations using ``LDA+U+ALPS''. The deviation at low temperature is due to impurities in the experimental sample, causing a Curie-Weiss like divergence at low temperatures. The dominant couplings are an intra-chain coupling $J=162$meV and an inter-chain coupling $J_\perp=76$meV}
\label{fig:srcu2o3}
\end{figure}

Release 1.4 of ALPS will contain tools to automate such calculations, based on a standard XML I/O format for band structure codes developed by T. Schulthess and M. Summers at the Oak Ridge National Laboratory. As an example of such an automated calculation using a prototype tool by A. Kozhevnikov, we show in figure \ref{fig:srcu2o3} a comparison of experimental measurements on the spin ladder compound SrCu$_2$O$_3$ \cite{azuma} with automated ab-initio simulations (``LDA+U+ALPS'').

\subsection{Dynamical mean field theory (DMFT) solvers}

In a further stage, we plan to expand the ALPS project to include dynamical mean field theory (DMFT) \cite{dmft} solvers. We have designed a DMFT framework for QMC solvers and implemented the standard Hirsch-Fye solver \cite{hirschfye}, as well as two new continuous time solvers \cite{rubtsov,Werner06}. The first application of the framework, a careful comparison of the performance of the three solvers \cite{hirschfye,rubtsov,Werner06} is in preparation \cite{compare}. Work is in progress to expand the framework to include ED and DMRG based solvers, and we expect a public open-source release of the first codes soon. Interested researchers can contact us to participate in the development or to obtain pre-release versions of the codes.

\section{Conclusions}

%The ALPS project is an open source effort to provide libraries and
%applications for the simulation of classical and quantum lattice
%models.  The
%ALPS applications make modern high-performance numerical algorithms
%available to a wider range of researchers. They enable theoreticians
%to investigate properties of interesting strongly correlated models
%conveniently, and allow experimentalists to perform direct comparisons
%and fits of experimental measurements to numerical simulations.

The ALPS project is continuously evolving. Researchers interested in announcements of new releases, information
about workshops, or in contributing to the ALPS project are encouraged
to sign up to the mailing lists on our web page \cite{alps}.

We acknowledge support by the Swiss National Science Foundation, the
Kavli Institute for Theoretical Physics at the University of
California at Santa Barbara, the Aspen Center for Physics, CNPq (Brazil), the Network for International Development and Cooperation (NIDECO) at ETH Z\"urich, the RGC grant of Hong Kong, and the Deutsche Forschungsgemeinschaft through SFB 602.


\begin{thebibliography}{99}
\bibitem{contact} Please  contact the authors if you need this or another feature.
\end{thebibliography}
\end{document}